\documentclass[twocolumn]{article}
\usepackage{graphicx}
\usepackage{subcaption}
\usepackage{lipsum}
\usepackage{amsmath,amsthm,amssymb, bm}
\usepackage{physics}
\usepackage{braket}
\usepackage{mathtext}
\usepackage{bm}
\usepackage{latexsym}
\usepackage[colorlinks=true, pdfborder={0 0 0}, linkcolor=red]{hyperref}
\usepackage{caption}

\begin{document}

\twocolumn[
\begin{center}

\textbf{\Large
Towards  out of equilibrium impurity solver for the \\ Dynamical Mean Field Theory}
\vspace{2mm}

{

\textbf{A. Karpov$^{1,2}$, G.Sultanov$^{1,3}$, E. Polyakov$^1$, and A. Rubtsov$^{1,2}$}

\textit{$^1$Russian Quantum Center, 30 Bolshoy Boulevard, building 1, Skolkovo Innovation Center territory, Moscow, 121205, Russia }\\
\textit{$^2$Department of Physics, Lomonosov Moskow State University, Leninskie gory 1, 119991 Moscow, Russia}
\textit{$^3$National Research Nuclear University MEPhI, Moscow, 115409, Russia}

e-mail: \textit{duha.kar.03@mail.ru}

}

\vspace{2mm}
\textbf{\large Abstract}

The Dynamical Mean Field Theory (DMFT) is a powerful tool for calculating highly correlated systems (both bosonic and fermionic) in a state of thermodynamic equilibrium. However, in the case of non-equilibrium states, the method has significant limitations that do not allow obtaining correct results. The stumbling block here is the impurity solver: a method for calculating the dynamics of an open system.
In this work we present the prototype of a universal impurity solver for the DMFT method which can solve both equilibrium and non-equilibrium problems. We analyse spectral functions of Bose-Hubbard model on the Bethe lattice in different regimes and show that our solver gives correct and physical results. 

\end{center}
]

\thispagestyle{empty}

\section{Introduction}
The development of methods such as pump-probe spectroscopy has opened up opportunities for studying time evolution on the scale of electronic motions \cite{Wall_2010}\cite{Sentef_2013}. By exciting the system with ultrashort pulses and then probing with a strobing pulse, we can observe the relaxation process to thermodynamic equilibrium. While quantum systems in thermodynamic equilibrium are well described by statistical mechanics, the description of nonequilibrium systems faces many technical and fundamental problems.

The process of transition of the system to a stable state after external excitation is a non-equilibrium process, otherwise called thermalization.
Most systems are thermalized, and in the process of relaxation, they lose information about their initial state; however, there are such phenomena and states of matter for which this principle is violated. For example, these are systems with many body localization\cite{Abanin_2019}, anomalous Floquet insulators\cite{Rudner_2019}, anomalous non-thermal states in Rydberg atoms\cite{Serbyn_2021} and much more.

The study of such phenomena is especially relevant today. 
Firstly, it raises fundamental questions in thermodynamics, for example, about how microscopic dynamics lead to thermodynamic equilibrium. 
Secondly, the study of systems like MBL can open possibilities for creating states resistant to decoherence, which is especially important in quantum computing. 
Therefore, numerical approaches to modeling non-equilibrium dynamics are actively developing now. One of these is the non-equilibrium dynamical mean field theory\cite{Aoki_2014}.

The dynamical mean field theory (DMFT) \cite{Kotliar_2004}\cite{Georges_96} is the famous tool in the study of strongly correlated fermions or bosons. 
The key idea of this method is that we map our many body system to the problem of interaction of a quantum impurity with an external self-consistent environment. The most significant part here is an impurity solver that is responsible for solving the impurity problem. Such problem is essentially an open system task. 
Modern solvers are well suited for fermionic systems in equilibrium \cite{Kotliar_2004}. Equilibrium bosonic systems are also solvable but require modifications of the classical DMFT cycle \cite{Byczuk_2008}.
In the non-equilibrium case significant limitation arise and this area is now developing.
\begin{figure}[h!]
\center{\includegraphics[width=0.9\linewidth]{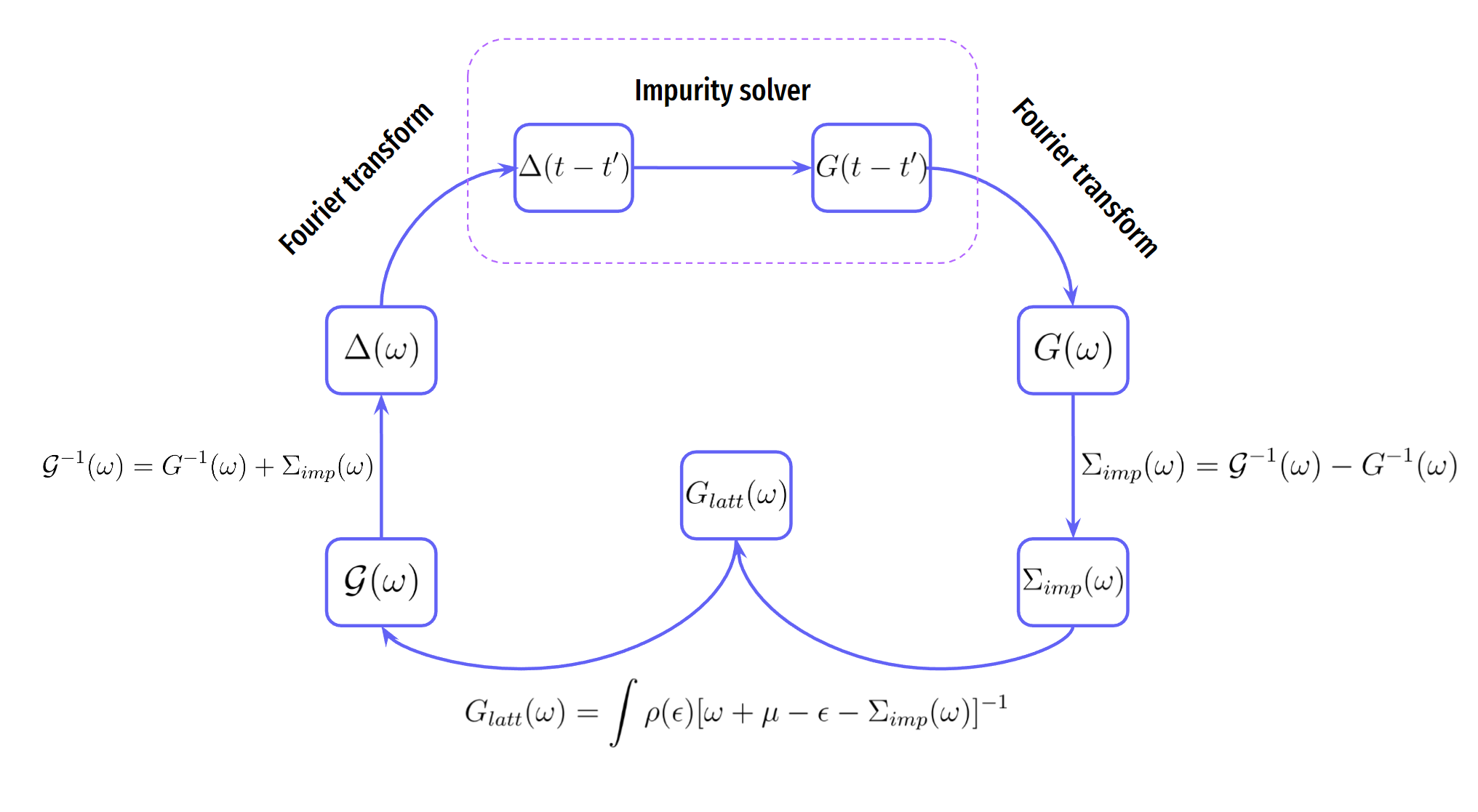}}
\label{fig:DMFT}
\caption{The conception of DMFT}
\end{figure}

In this work we present the prototype of a universal impurity solver for equilibrium and non-equilibrium problems. 
It is based on a new approach to solving the non-stationary quench problems for open quantum systems on wide time scales \cite{Polyakov_2022}. 
In particular, it allows calculate DMFT cycle in real time without analytical continuation. 
Method reduces the complexity of calculations using the conception of the Lieb-Robinson bound (like a light cone) as a selection criteria of effective modes of the environment.  
In addition to the forward "Lieb-Robinson light cone", we can build a reverse light cone due to the time reversal symmetry of the Schrodinger equation.  The intersection of such light cones gives us the criterion  for cutting off irrelevant modes.

We check the correctness of our solver by the example of calculating the spectral functions of the Bose-Hubbard model on the Bethe lattice
and show that our solver gives physical results. 
In particular, we calculate correlation functions over a long time range and confirm, that for the non-interacting regime ($U=0$) the results coincide with the known analytical solution.


\section{The method of approximate estimation of the light cones}
\subsection{Main idea}
In this paper we use method which was proposed in works  \cite{Polyakov_2022}\cite{Polyakov_jan_2022}. 
 
We propose the physical criterion, that can limit the number of modes of the environment in such a way that it remains fixed throughout the time evolution of the system.
As a result, it turns out that the open system interacts only with effective mods, and not with the full environment.
This makes it possible to overcome the difficulties associated with the exponential growth of the Hilbert space when calculating non-equilibrium dynamics over long periods of time.

In its ideological essence, the light cone method is a renormalization group. Examples of such methods are NRG\cite{Wilson_1975}\cite{Bulla_2008} and DMRG\cite{White_1993}\cite{Kumar_2016}. 

Such approach allows us to find such a mapping of the original Hamiltonian (or action), after which it is possible to distinguish the essential and non-essential degrees of freedom (states) necessary to describe the system. For example, as mentioned above, in the NRG method, states with high energy are cut off, which allows you to keep the dimension of the Hilbert space constant when adding a new node to the chain. In analogy, in equilibrium thermodynamics, the canonical Gibbs density matrix $\hat\rho = \exp(-\beta\hat H)$ \textit{a priori} sets statistically significant energy levels due to finite temperature.
The lightcone method also cuts off irrelevant states, but in the case of the time evolution. Let's take a closer look at this process.

Let us consider a system of the following form:
\begin{equation}
    \hat H = \hat H_0 + \hat H_{bath} + \hat H_{int},
\end{equation}
where  $\hat H_0$, $\hat H_{bath}$ describe the subsystem and the environment, respectively, and $\hat H_{int}$ is the Hamiltonian of its interaction. Let us assume continuous spectrum of the bath:
\begin{equation}
    \hat H_{bath} = \int_0^{\infty} dw w \hat b^{\dag}(w) \hat b(w),
    \label{eq:H_E}
\end{equation}
In this case, let the subsystem be a single mode:
\begin{equation}
    \hat H_0 = \varepsilon_0 \hat b^{\dag}_0 \hat b_0,
    \label{eq:H_S}
\end{equation}
and its interaction is represented in the form of:
\begin{equation}
    \hat H_{int} = \hat b_0 * \int_0^{\infty} dw c(w) \hat b(w) + \hat b_0^{\dag} * \int_0^{\infty} dw c(w) \hat b^{\dag}(w).
    \label{eq:H_int}
\end{equation}
Representation in the form of (\ref{eq:H_E}), (\ref{eq:H_S}) and (\ref{eq:H_int}) is the so-called representation of an open system in the form of a "star". From this representation, after discretization, we can proceed to the representation of the Hamiltonian of the system in the form of a chain \cite{Chin_2010}, where the zero node corresponds to the impurity:
\begin{equation}
\begin{split}
    \hat H = \varepsilon_0 \hat b^{\dag}_0 \hat b_0 + (h_0 \hat b_0^{\dag} \hat{b}_1 + c.c.) + \\
    +\underbrace{\sum^{\infty}_{k=1} h_k(\hat{b}^{\dag}_{k} \hat{b}_{k+1} + c.c.) + \sum^{\infty}_{k=1} \varepsilon_k \hat{b}^{\dag}_k \hat{b}_k}_{\hat H_{bath}}.
    \label{eq:H_chain}
\end{split}
\end{equation}

In the interaction picture, we can exclude the environment $\hat H_{bath}$ from the full Hamiltonian, while:
\begin{equation}
    \hat H(t) = e^{it \hat H_{bath}} \hat H e^{-it \hat H_{bath}} - \hat H_{bath},
    \label{eq:int_pic}
\end{equation}
\begin{equation}
    \hat b_1(t) = e^{i t \hat H_{bath}} \hat b_1 e^{-i t  \hat H_{bath}}.
\end{equation}
In the resulting picture, the external environment is a continuum of free modes, and the interaction of $\hat H_{int}$ depends on time. In such a picture, we can write the evolution of the impurity node as:
\begin{equation}
    \hat b_0(t) = \sum_{k=0}^{\infty} \phi_k(t) \hat b_k,
    \label{eq:modes}
\end{equation}
where $\phi_k(t)$ is the solution of the primary quantized Schrodinger equation:
\begin{equation}
    \begin{cases}
    i\frac{\partial \phi_k(t)}{\partial t} = \hat H_{bath} \phi_k(t)\\
    \phi_k(0) = \delta_{k,1}
    \end{cases}
\end{equation}
\begin{equation}
    \begin{split}
        \hat H_{bath} = \left[\begin{array}{cccc}
        \varepsilon_1 & h_1 & 0 & \dots\\
        h_1 & \varepsilon_2 & h_2 & \dots\\
        0 & h_2 & \varepsilon_3 & \dots\\
        \vdots &  & \vdots & \ddots
        \end{array}\right]
    \end{split}
\end{equation}

In the sense of $\phi_k(t)$ describes the interaction of the k-th mode with an impurity.
It is clear from physical considerations that a single-particle impurity cannot interact with all the modes of the environment at the initial moment (for example, at the initial moments of time, distant nodes in the chain "don't feel" the impurity).

Thus, over time, the impurity node will interact with even more modes, which are gradually "connected" to the system (incoming modes):
\begin{equation}
    \hat b_0(t) = \sum_{k=0}^{\infty} \phi_k(t) \hat b_k \longrightarrow \sum_{k=0}^{m(t)} \phi_k(t) \hat b_k,
\end{equation}
where $m(t)$ is a monotonically growing function. In terms of meaning, this is just connected with the idea of the Lieb-Robinson boundary\cite{SIMS_2011} or with the concept of a light cone.

\subsection{Out-of-Time-Order Correlator}
How can we evaluate the function $m(t)$ \textit{a-priori} before solving a many body problem? The criterion for selecting relevant modes at time $t$ can be \textit{Out-of-Time-Order Correlator or OTOC)}\cite{Roberts_2016}:
\begin{equation}
    C_k(t) = \|[\hat b_0(t), \hat b_k^{\dag}]\| = \sqrt{\bra{0}[\hat b_0(t), \hat b_k^{\dag}] [\hat b_0(t), \hat b_k^{\dag}]^{\dag}\ket{0}},
\end{equation}
where $\bra{0} ... \ket{0}$ -- averaging over the vacuum state (note that for a linear environment, the commutator is a number, so the averaging can be taken over an arbitrary state). OTOC shows the instantaneous intensity of the interaction of the k-th mode with the impurity at time $t$. The equality of OTOC to zero means that at time $t$ the k-th mode is not coupled with an impurity.

We can consider the intensity not only for modes in the original lattice basis, but also for their linear combination:
\begin{equation}
    \hat\kappa = \sum_{k=0}^{\infty} \chi_{k} \hat b_k.
    \label{eq:kappa}
\end{equation}

In this case, using (\ref{eq:modes}) and (\ref{eq:kappa}), it is not difficult to make sure that:
\begin{equation}
\begin{split}
    [\hat b_0(t), \hat\kappa^{\dag}] = [\sum_{k=0}^{\infty} \phi_k(t) \hat b_k ,\sum_{l=0}^{\infty} \chi_{l}^* \hat b_l^{\dag}] =\\= \sum_{k=0}^{\infty} \chi_{k}^* \phi_k(t) =
    \braket{\chi | \phi(t)},
\end{split}
\end{equation}
where $\bra{\chi}$ is a column vector composed of coefficients $\chi^{*}_k$.
Then OTOC can be written as:
\begin{equation}
    C(t, \chi) = \braket{\chi | \phi(t)} \braket{\phi(t) | \chi}.
\end{equation}
\subsection{Forward lightcone}
Since the OTOC shows the coupling of the impurity with the external mode at time $t$, and we are interested in how this mode interacts with the impurity in the time interval $[0,t]$, we take the average value of the OTOC in this interval:
\begin{equation}
\begin{split}
    \int_0^{t} d\tau C(\tau, \chi) = \bra{\chi} \underbrace{\int_0^{t} d\tau \ket{\phi(\tau)} \bra{\phi(\tau)}}_{=\hat\rho_{+}} \ket{\chi}= \\ = \bra{\chi} \hat\rho_{+}(t) \ket{\chi}.
\end{split}
\end{equation}
It turns out that if the mode interacts negligibly little with the impurity in the time interval, then the average value of the OTOC will be small. Physically, this means that the mode is located beyond the boundary of the light cone.
To accurately determine this boundary, for the resulting density matrix $\hat\rho_{+}$, we introduce a significance threshold $a_{cut}$:
\begin{equation}
    g(t,\chi) =  \bra{\chi} \hat\rho_{+}(t) \ket{\chi} - a_{cut}.
    \label{eq:g_cut}
\end{equation}
Thus, we can assume that the $\hat\kappa$ mode is inside the light cone if:
\begin{equation}
    g(t,\chi) \geqslant 0.
    \label{eq:g_criteria}
\end{equation}
For convenience, we will call the resulting metric the ``Lieb-Robinson metric''.

Using the criterion (\ref{eq:g_criteria}) at each time interval, we can determine the modes with which interaction will be significant. With time evolution, there is a gradual "picking up"\ such modes, and with the increase in their number $m_{in}(t)$, the dimension of the Hilbert space increases. This is exactly the stumbling block of most methods for calculating the dynamics of open systems (for example, the method of exact diagonalization).

\subsection{Backward light cone}
Let the impurity subsystem evolve on the interval $[0,T]$.
We can construct the Lieb-Robinson metric not only for the initial lattice basis, but also for an arbitrary:
\begin{equation}
    \hat\kappa_i = \sum_{k=0}^{\infty} \chi_{ik} \hat a_k.
    \label{eq:kappa}
\end{equation}
It is clear that each basis will have its own rate of picking up modes. Does the question arise whether it is possible to find such a basis in which this speed will be minimal? It turns out that yes, so later in the paper we consider the light cone in this basis. The procedure for finding it is described in detail in \cite{Polyakov_2022}.

In order to estimate the inverse light cone a priori, by analogy we also introduce OTOC and average it over future times $[t, T]$:
\begin{equation}
\begin{split}
    \int_t^{T} d\tau C(\tau, \chi) = \bra{\chi}  \underbrace{\int_t^{T} d\tau\ket{\phi(\tau)} \bra{\phi(\tau)}}_{=\hat\rho_-} \ket{\chi} =\\= \bra{\chi} \hat\rho_{-}(t) \ket{\chi}.
\end{split}
\end{equation}
Using the obtained matrix elements $\hat\rho_{-}$, we construct the same significance criterion (\ref{eq:g_criteria_minus}), only for "uncoupling"\ modes. Their number $m_{out}(t)$ will also grow over time.

It is important to note that the matrix elements $\bra{\chi} \hat\rho_{-}(t)\ket{\chi}$ we construct $m_{in}$ in the space of incoming modes. Thus using the criterion:
\begin{equation}
g(t,\chi^{(in)}) = \bra{\chi^{(in)}} \hat\rho_{-}(t) \ket{\chi^{(in)}} - a_{cut} \geqslant 0
\label{eq:g_criteria_minus}
\end{equation}
you can "cut off"\ irrelevant mods among the incoming ones at each time interval. This corresponds to a situation when at a certain moment the imcoming mode ceases to significantly affect the impurity node and can be considered as independent.

\section{Model}
In this paper we solve the Bose-Hubbard model on the Bethe lattice using DMFT and lightcone method. The initial Hamiltonian
of the system has the form:
\begin{equation}
    \hat H = -t\sum_{<i,j>} (\hat b^{\dag}_{i} \hat b_{j} + \hat b^{\dag}_{j} \hat b_{i}) + 
    \mu \sum_{i} \hat n_{i} + 
    \frac{U}{2} \sum_i \hat n_i(\hat n_i-1),
\end{equation}
where $\hat b^{\dag}_i$ and $\hat b_i$ are the boson creation and annihilation operators on site $i$, respectively. $\hat n_{i}$ is the boson number operatoron the site $i$. Here, we consider the hopping amplitude $t$, the chemical potential $\mu$ and on-site energy $U$.

The initial approximation of the local Green function
is proposed to be set using the density of states for
non-interacting bosons with the number of nearest neighbors tending
$Z$ of each node to infinity:
\begin{equation}
    \rho_0(x) = 
    \begin{cases}
    \frac{\sqrt{4t^2-x^2}}{2\pi t^2}&, x \leqslant 2t \\
    0 &, x > 2t
    \end{cases}
    \label{dos_bethe}
\end{equation}
according to the equation:
\begin{equation}
    G^{(0)}(\omega) = \int_{0}^{\infty} dx \frac{\rho_0(x)}{\omega +\mu - x}.
\end{equation}

To use the DMFT method, it is necessary to map the original problem to an open system model:

\begin{equation}
    \hat H_{imp} = \hat H_{0} + \hat H_{hyb} + \hat H_{bath}
\end{equation}

Let $\hat b_0^{\dag} (\hat b_0)$ be the impurity creation(annihilation) operators, $\hat b^{\dag}(w) (\hat b(w))$ -- operators of the creation(annihilation) of the mode $w$.
Then the resulting system can be rewritten as:

\begin{equation}
     \hat H_{imp} = \hat H_{0} + (\hat b_0^{\dag} \hat V + c.c.) + \int dw w \hat b^{\dag}(w) \hat b(w),
\end{equation}
where $\hat V = \int dw c(w) b(w)$ with $c(w)$ -- the coupling between bath mode $w$ and impurity mode. The impurity Hamiltonian:
\begin{equation}
    \hat H_{0}  = -\mu \hat n_0 + \frac{U}{2} \hat n_0 (\hat n_0 - 1),
    \label{eq:H_0}
\end{equation}
where $\hat n_0 = \hat b^{\dag}_0 \hat b_0$.

The resulting system has a "star" topology where the impurity is coupled to the surrounding mods.
It allows us to map such a system to the chain \cite{Chin_2010} where the zero node is responsible for the impurity (\ref{eq:H_chain}).
To proceed to such a statement of the problem, we perform
a transformation of the modes $\hat b_i$ by reducing the Hamiltonian to a tridiagonal form like 

The impurity node will gradually becomes entangled with the nodes of the chain over time, and we have the same situation that we considered in the previous section. 
The process of entanglement has a finite propagation velocity, which is limited by the Lieb-Robinson bound. 
In every time moment, the lightcone method allows us to extract relevant modes using criteria (\ref{eq:g_criteria}) and (\ref{eq:g_criteria_minus}).

\section{DMFT on the Bethe lattice}
In the case of a Bethe lattice in the continuum limit of the number of connections ($Z\longrightarrow\infty$), the DMFT cycle can be greatly simplified. The main feature is the fact that the conditions for matching the DMFT equations take on a simple form:
\begin{equation}
    \Delta(\omega) = t^2 G(\omega).
    \label{eq:delta}
\end{equation}

As a result, the entire DMFT cycle is reduced to a sequence of the following operations:

0. Set the initial approximation for the local Green function $G(\omega)$. For example, a Green function for a non-interacting system ($U=0$);

1. Calculate the hybridization of $\Delta(\omega)$ by the formula (\ref{eq:delta});

2. Solve the impurity problem and calculate $G_{imp}(\omega)$;

3. $G = G_{imp}$.

4. Repeat steps 1-3 until convergence.

\subsection{Solution of the impurity problem}
The mapping to the solution of the impurity problem (point 2 of the DMFT cycle) is given in Section 2.1. From the hybridization function $\Delta$ we can find the coefficients $c(w)$ by formulae:
\begin{equation}
    c(w) = \pi \sqrt{\Delta(w)}.
    \label{c_w_i}
\end{equation}
After that, mapping the system to a linear chain, we come to the Hamiltonian (\ref{eq:H_chain}). In the interaction representation (\ref{eq:int_pic}), the resulting Hamiltonian is written as:
\begin{equation}
    \hat H_{imp}(t) = \hat H_0 + (h_0 \hat b^{\dag}_0 \hat b_1 + c.c.),
\end{equation}
where
\begin{equation}
    \hat b_1(t) = e^{i t \hat H_{bath}} \hat b_1 e^{-i t  \hat H_{bath}}.
\end{equation}
Next, we assume that the environment is in thermodynamic equilibrium in the state $\frac{1}{Z}e^{-\beta\hat H_{bath}}$. Here $\beta =\frac{1}{T}$ characterises the ambient temperature. At time $t=0$, we add an impurity to the original environment. The general state of the system at this moment is written as:
\begin{equation}
    \hat \rho_0(0) = \ket{0}\bra{0}_{imp} \otimes \frac{1}{Z} e^{-\beta \hat H_{bath}}.
\end{equation}

The light cone method calculates the evolution of this state over time:
\begin{equation}
    \hat \rho_0(t) = U(t) \hat \rho_0(0) U^{\dag}(t),
\end{equation}
where $U(t)=\mathcal{T} e^{-i\int_0^{t}d\tau \hat H_{imp}(\tau)}$, $\mathcal{T}$--time ordering.

Let $t_{0}$ be the time to establish equilibrium in the system after adding an impurity. As a result, we have:
\begin{equation}
    G_{imp}(t) = -i(\langle\hat b_0(t+t_{0}) \hat b^{\dag}_0(t_{0})\rangle - \langle\hat b^{\dag}_0(t+t_{0}) \hat b_0(t_{0})\rangle),
\end{equation}
where averages:
\begin{equation}
\begin{split}
    \langle\hat b_0(t+t_{0}) \hat b^{\dag}_0(t_{0})\rangle = Tr\{\hat b_0 U(t+t_0) \hat b^{\dag}_0 \hat\rho_0(0) U^{\dag}(t+t_0)\},
\end{split}
\end{equation}
\begin{equation}
    \langle \hat b^{\dag}_0(t+t_{0})\hat b_0(t_{0})\rangle = Tr\{\hat b^{\dag}_0 U(t+t_0) \hat b_0 \hat\rho_0(0) U^{\dag}(t+t_0)\}
\end{equation}
 are calculated using light cone method.

\section{Results}
\subsection{Non-interacting limit $U=0$}
The Hubbard model on the Bethe lattice in the absence of interaction ($U=0$) between particles (fermions or bosons) have exact solution. The density of the energy states of such a system is given by the formula (\ref{dos_bethe}).

Another important feature is that in the context of the DMFT method, the desired solution is found in a small number of iterations (typically 5-10).

\begin{figure}[h!]
    \center{\includegraphics[width=0.9\linewidth]{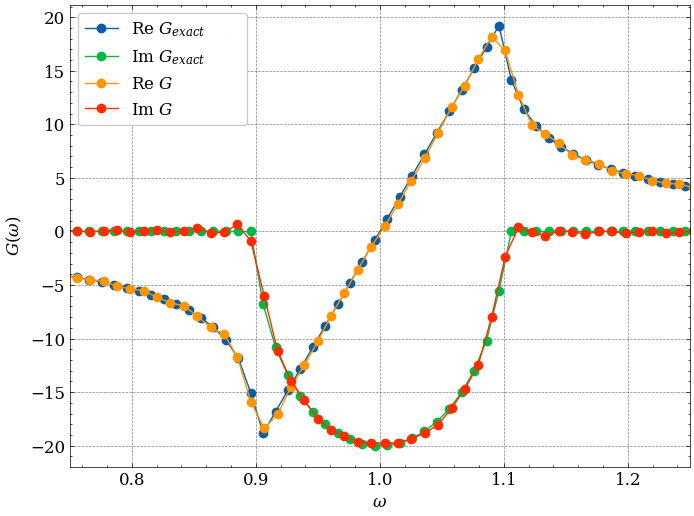}}
    \label{fig:U=0}
    \caption{The Green function of the Bose-Hubbard model on the Beta lattice at $U=0$ with parameters $t=0.05$, $\beta=10$, $\mu=1$}
\end{figure}

In Fig.2 
shows the exact Green function and the Green function obtained by the DMFT method using the lightcone solver. Since we choose the initial approximation in the form of the Green function of a non-interacting system, the desired solution is found quite quickly. 

Non-interacting limit also allows us to clearly show the convergence of the solution by the number of modes of the environment in the impurity solver (see Fig.3 
). The convergence turns out to be quite fast and at 5-6 modes the resulting solution is already slightly different from the exact one. We use such parameters in further calculations.

\begin{figure}[h!]
    \center{\includegraphics[width=0.9\linewidth]{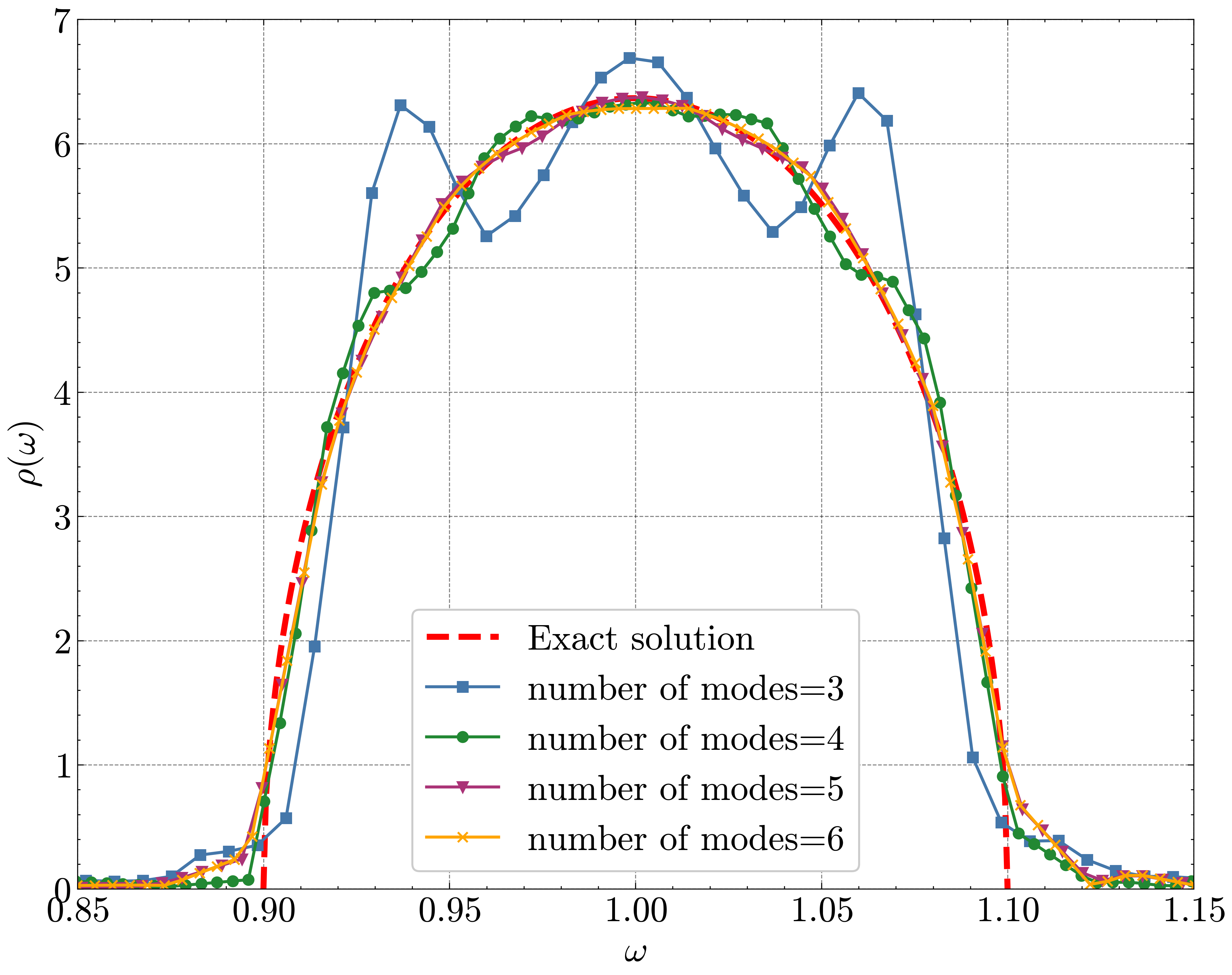}}
    \label{fig:modes}
    \caption{Convergence of the method by the number
of relevant modes of the environment $r$}
\end{figure}

\subsection{Atomic limit $\frac{t}{U} << 1$}
The limit $t=0$ describes a situation where particles remain localized at nodes. In this mode, at low temperatures, the system is in the Mott insulator phase: the number of particles at the node becomes fixed. The density function of states in this case takes the form of delta functions at the corresponding energies. At zero temperature, in the case of an increase in $t$ in the Bose-Hubbard model, a phase transition occurs between the Mott phase and the superfluid phase (where condensate cannot be neglected) \cite{Hu_2009}. Formally, an increase in $t$ leads to an increase in the width of the energy peaks, which leads to their overlap in the density of states.

\begin{figure}[h!]
    \center{\includegraphics[width=0.9\linewidth]{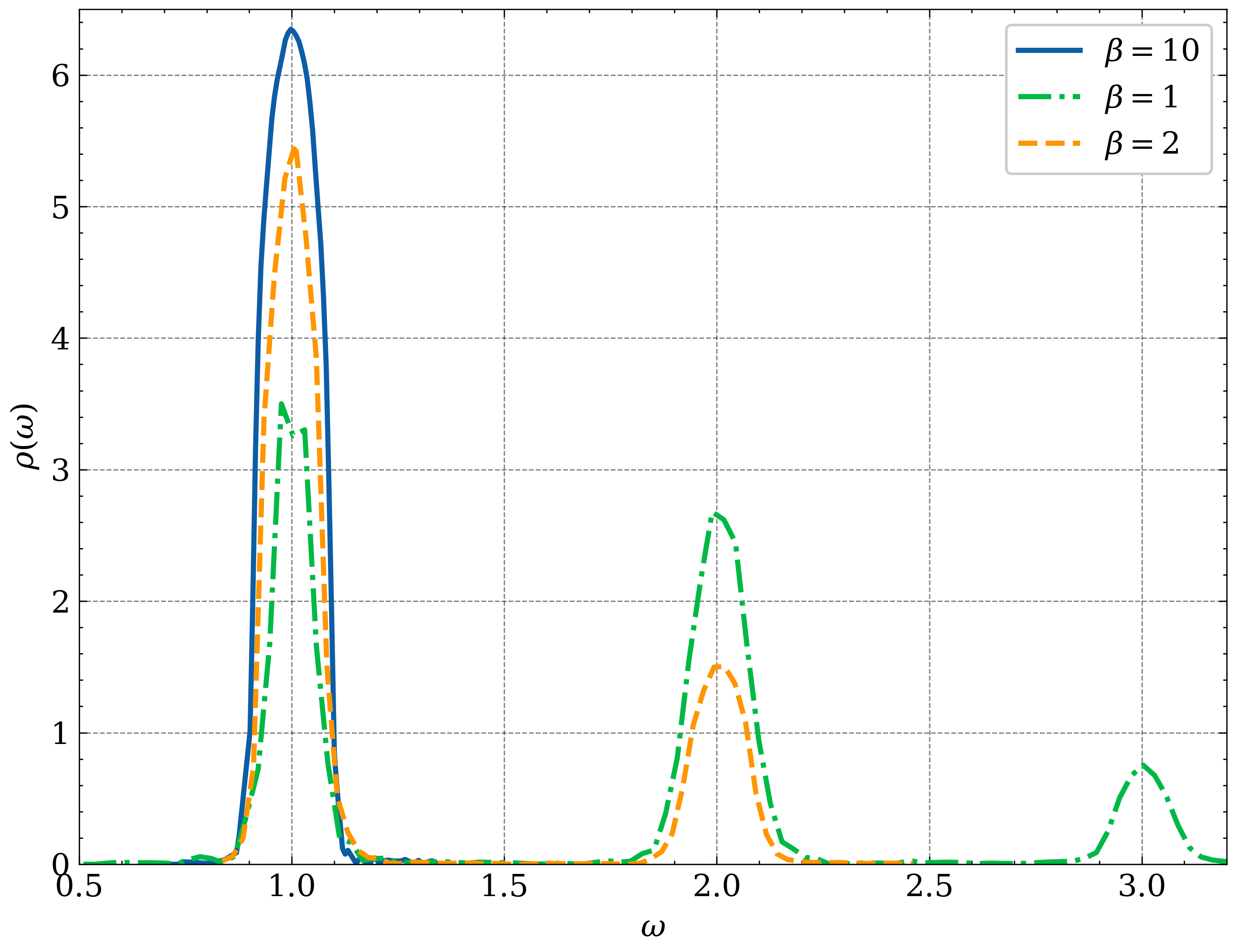}}
    \label{fig:t=0.05}
    \caption{Temperature dependence of the spectral density \newline at $t=0.05$, $U=1$, $\mu=1$}
\end{figure}

Since we consider the system in equilibrium, the parameters $\mu$ and $\beta$ set the available number of energy levels. For large values of $\beta$ (small $T$), according to the Boltzmann distribution, the presence of two or more particles on the node may be low probability, as a result of which the term $\frac{U}{2}n(n-1)$ does not contribute to the energy levels of the system and we observe same picture like in non-interacting limit $U=0$.

With a decrease in $\beta$ (large $T$), the number of levels available for occupation increases, and the probability of finding two or more particles at the node increases. Fig. 4 
shows that the number of peaks increases with increasing temperature. In this case, the distance between them is determined by the parameter $U$ (in our case, it is equal to 1).

\subsection{The case of arbitrary $t$}
\begin{figure}[h!]
  \begin{subfigure}[b]{0.9\columnwidth}
    \includegraphics[width=\linewidth]{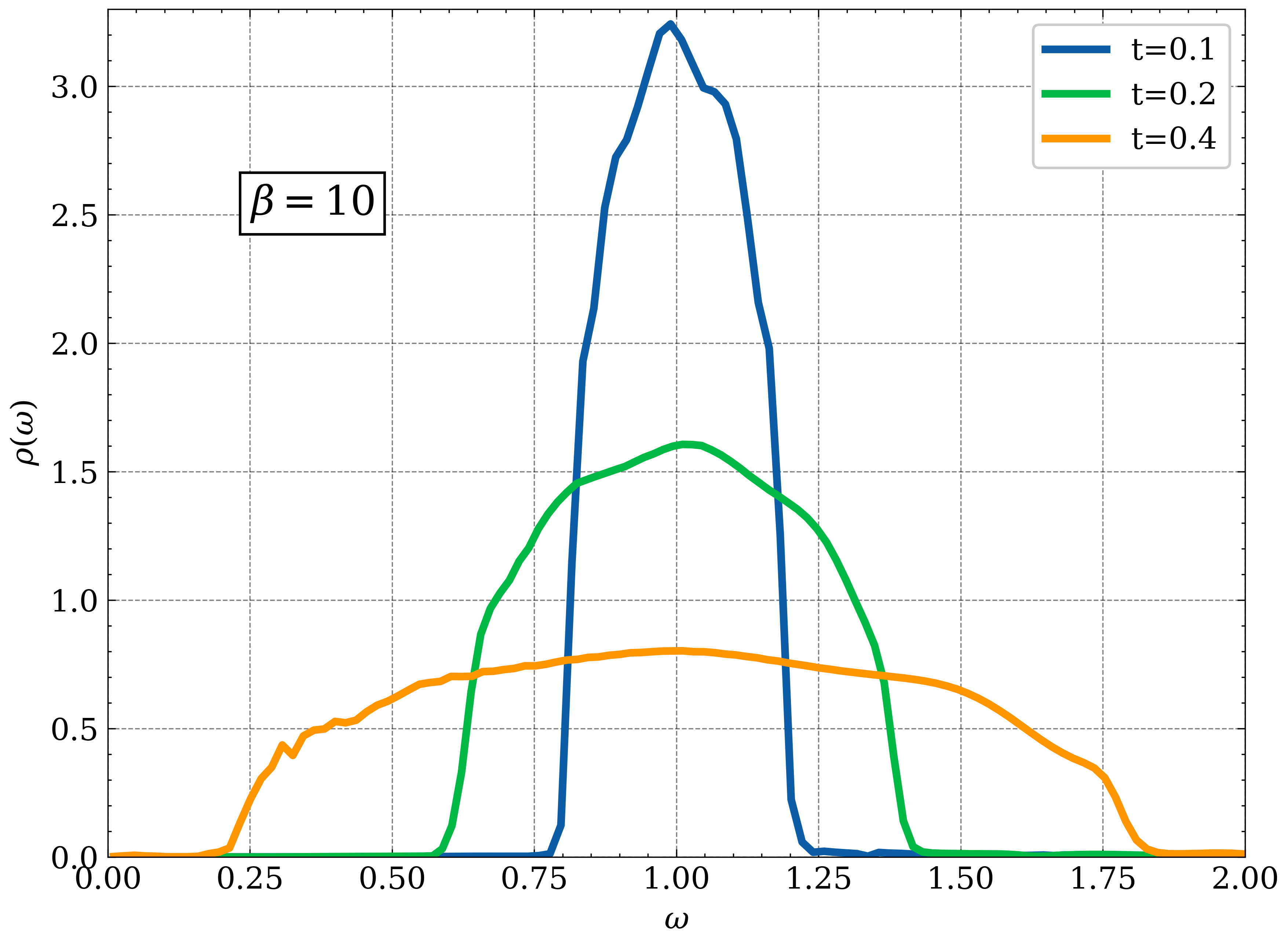}
    \caption{$\beta = 10$}
    \label{fig:dependence_t_beta=10}
  \end{subfigure}
  \hfill 
  \begin{subfigure}[b]{0.9\columnwidth}
    \includegraphics[width=\linewidth]{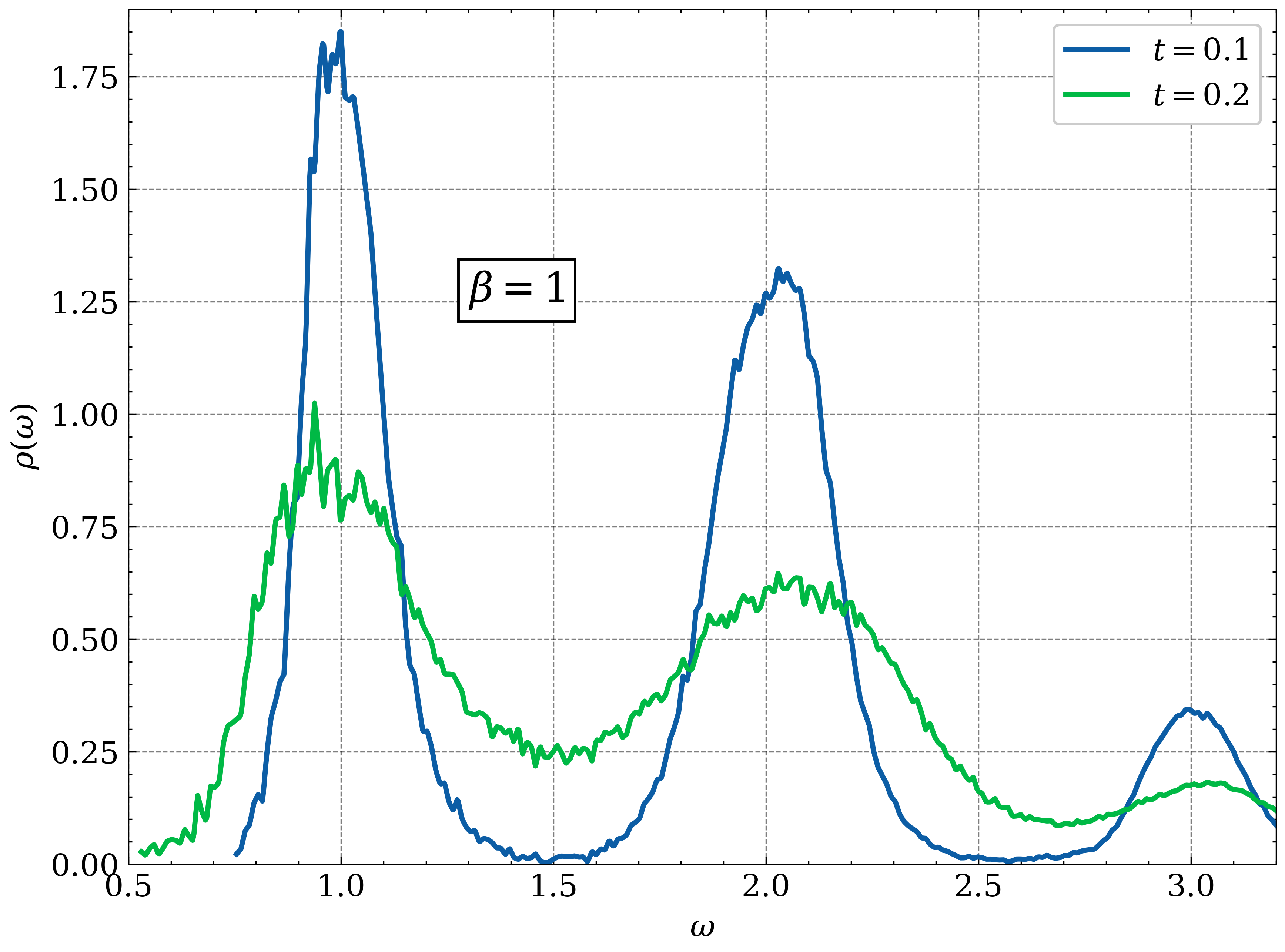}
    \caption{$\beta = 1$}
    \label{fig:dependence_t_beta=1}
  \end{subfigure}
\caption{Dependence of the spectral density on the value of the hopping $t$ at the temperature: 1) $\beta=10$, 2) $\beta=1$}
\label{fig:figures}
\end{figure}

As the value of the hopping $t$ increases, the width of the peaks that arise increases, as a result of which the mobility of bosons in the lattice increases. This is most clearly seen in Fig.\ref{fig:dependence_t_beta=10}
In the case of low temperatures, the excitations of high-energy levels are limited by the Boltzmann distribution. Therefore, we observe a broadening of only one peak in the spectral density. In the case of high temperatures, as can be seen in Fig.\ref{fig:dependence_t_beta=1}, peaks separated from each other at a distance of $U=1$ merge, forming a complex shape of spectral density.

The noise that occurs in the case of high temperatures in the results is related only to the features of numerical calculation in the light cone method and has no physical nature.

\section{Conclusion}
In this paper, we implemented an impurity solver in the method
of dynamic mean field theory based on a new approach to solving
non-equilibrium problems in open systems. Using a criterion
based on forward and backward light cones, the method allows
us to identify a limited number of modes with which the impurity interacts.
A significant advantage of the calculations performed is that
they are performed on the real-time axis, therefore they do not require an inaccurate analytical continuation procedure.

For the Bose-Hubbard model on a Bete lattice in non-interacting limit we were able to obtain results that coincided with the exact solution. Also we have considered the dependences of the obtained correlation functions on
temperature, the magnitude of the hoppings $t$. In the case of small $t$, i.e. in the atomic limit, the results coincide with the expected behavior of the system.

For all parameters, the method gives iterative convergence to
the result and show stability to initial conditions, while
the integral over the frequencies of the resulting spectral density remains equal to one.

In our future plans, we will continue to develop this work
and make calculations for other phases of the Bose-Hubbard model. Our task is to observe the transition between the phase with the presence of Bose condensation and the normal phase. Also in our plans to continue develop a method for fermionic models and implement a universal solver of a non-equilibrium impurity problem.



\bibliographystyle{unsrt}
\bibliography{main}

\end{document}